# MEASURING SCIENCE: IRRESISTIBLE TEMPTATIONS, EASY SHORTCUTS AND DANGEROUS CONSEQUENCES[1]

**Summary**

*In benchmarking international research, although publication and citation analyses should not be used to compare different disciplines, scientometrists frequently fail to resist the temptation to present rankings based on total publications and citations. Such measures are affected by significant distortions, due to the uneven fertility across scientific disciplines and the dishomogeneity of scientific specialisation among nations and universities. In this paper, we provide an indication of the extent of the distortions when comparative bibliometric analyses fail to recognise the range of levels of scientific fertility, not only within a given major disciplinary area but also within the different scientific disciplines encompassed by the same area.*

Governments, businesses and trusts charged with the responsibility of deciding on scientific priorities and funding are asking increasingly exacting questions of scientometrists in relation to the relative positioning of a given country or research organisation in terms of the quantity, quality and efficiency of the scientific research being carried out within it. Databases that provide complex and reliable information on the output of the research activities and on the relevant factors of production are widely available, and are proving so tempting that is it practically impossible not to get drawn into comparative evaluation exercises. The rankings that result from these exercises perfectly satisfy the innate human need to benchmark and also provide support for decision-making on the efficient allocation of resources. Lured in by the correspondence between the rankings and one's own expectations and/or by the use that can be made of the rankings, it becomes difficult to take due account of the reliability and the margin of error of the measuring system deployed. In certain cases, it is the scientometrists themselves who take shortcuts – as easy as they are risky – in order to satisfy the (to a greater or lesser extent latent) requirements of the demand side. But what consequences could result from inaccurate information on the scientific performance of this or that nation or research organisation.

In this article, we take on the unpopular duty of disappointing all those who – driven by needs that they feel to be essential – believe unquestioningly in these rankings. Moreover, and leaving aside professional deontology, we are sustained in our endeavours by the hope that we may contribute to the refining of methodologies of comparative bibliometric analysis and, in the process, help to avoid the dangerous consequences that could result from an imprudent use of the rankings.

In 2004, King[1] updated and extended the groundbreaking work done by May[2], providing metrics for evaluating the comparative performance in science and engineering research of various nations. To measure the quantity and quality of science in different countries, King analysed the numbers of published papers and their citations, provided by Thomson

---

Scientific, which indexes more than 8,000 journals – a figure that represents most of the significant material being published on science and engineering. The author flagged up few potential problems with this type of bibliometric analysis (a more extensive and in-depth analysis of technical and methodological problems affecting bibliometric measures may be found in Van Raan[3], which we strongly recommend to users). In particular, King warns that citation (and, we would add, publication) analyses should not be used to compare different disciplines. To do so is risky for two basic reasons: the first is that the plethora of scientific papers indexed by Thomson Scientific is unevenly distributed across various disciplines (for example, there are more papers on medical research than on mathematics, and so there are many more citations from medical research papers than from mathematics papers); the second is that the intensity of publication is not constant across the various fields of scientific research, since certain disciplines – due to their intrinsic characteristics – are more fertile than others[4]. This factor is a serious obstacle to aggregate performance comparisons, such as those focusing on countries or on multi-disciplinary research organisations, due to the inevitable distortions induced by the different levels of investment across different scientific disciplines. Nevertheless, in research papers, it is not unusual to find examples of bibliometric analyses for comparative measurements of the research performance of universities that disregard this limitation or that underestimate its distorting effects[5]. King himself, while underlining the aforementioned obstacle and providing evidence of the different national strengths in different disciplines (Fig. 3 of his paper), cannot resist the temptation to present international rankings of aggregate scientific performance (such as publications, citations, publications per researcher, citations per researcher, etc.; see Fig. 5 of his paper).

The ranking of national contributions to world science based on share of publications or citations is little more than an algebraic exercise, since the relevant information is either obvious (e.g. the fact that the USA has a larger scientific community than Sweden can be taken for granted) or subject to misleading errors (comparisons between nations/regions of comparable economic power, such as USA/EU or Sweden/Switzerland, can be distorted by not taking into account the relevant scientific specialisations).

What would happen if science policy scholars and policy decision-makers were to come up with research policy proposals based on performance rankings of this sort, leading the decision-makers, for example, to allocate research funds to one university over another? The purpose of this article is to provide an indication of the extent of the distortions to which aggregate measurements of research performance can be subject, highlighting the resultant risks and identifying a few regulatory implications.

The findings we will focus on are mostly the result of ongoing research into the assessment of the comparative scientific performance of Italian universities. Although the objective of the underlying research is admittedly parochial (at least in global terms), this paper is intended to be of more widespread interest, since it provides general evidence of the performance ranking distortions caused by crude disciplinary aggregations at both the output and input levels.

**Distortion assessment**

The main sources of information for our research were: i) the SCI[6] database of publications in science and engineering journals from Thomson Scientific, and ii) the MIUR[7] database of Italian universities' research personnel.

Each member of Italy's academic research staff is associated with a given Scientific Discipline (SD). There are 205 Science and Engineering Disciplines in total. These SDs are grouped into 9 major Disciplinary Areas (DAs). In total, there are 69 Italian universities that have research personnel in at least one of the science and engineering disciplines. By using a disambiguation algorithm and intervening manually, we managed to assign to each university author the relevant publications included in the SCI database during the period 2001-2003.

We then investigated the extent to which these publications were representative of the entire body of research output, calculating the number of scientific articles recorded in the SCI as a percentage of the total of all research products presented by the universities in the first Italian Research Assessment Exercise, relating to the same period. The percentage was higher than 90% in all of the DAs, with the exception of the DA: Civil Engineering and Architecture[8], which we then excluded from our subsequent analyses. We therefore have some considerable degree of confidence that our approach has been meaningful in terms of establishing a benchmark. Table 1 shows the rank orders of DAs based on research staff, publications[9], and publication intensity – PI (publications per researcher).

**Table 1 - Rank orders of Disciplinary Areas based on researchers, articles, and publication intensity, average 2001-2003**

| Disciplinary Area | Universities | Researchers | Rank | Publications | Rank | PI | Rank |
|---|---|---|---|---|---|---|---|
| 1 – Mathematical sciences | 59 | 3108 | 5 | 1011 | 6 | 0.33 | 8 |
| 2 – Physical sciences | 57 | 2516 | 7 | 2787 | 4 | 1.11 | 2 |
| 3 – Chemical sciences | 58 | 3150 | 4 | 4116 | 3 | 1.31 | 1 |
| 4 – Earth sciences | 48 | 1291 | 8 | 569 | 8 | 0.44 | 6 |
| 5 – Biological sciences | 63 | 4866 | 2 | 4257 | 2 | 0.87 | 3 |
| 6 – Medical sciences | 57 | 10571 | 1 | 7922 | 1 | 0.75 | 4 |
| 7 – Agricultural and veterinary sciences | 49 | 2964 | 6 | 1002 | 7 | 0.34 | 7 |
| 8 – Industrial and information engineering | 60 | 4350 | 3 | 2019 | 5 | 0.46 | 5 |

The data makes clear that the comparison of universities of a similar size, based on the total number of publications, in terms of the contribution made to the national knowledge base, is subject to major distortions due to the uneven scientific fertility across the various DAs. Unless, of course, the universities are considered as being perfectly homogeneous in terms of the resources available in the different areas, but to do so would be to adopt a position that lies some distance from reality, at least as far as Italy is concerned.

The dishomogeneity of scientific specialisation among nations as shown by the measurements conducted by Adams through an analysis of the relative citation impact[10], leads to the conclusion that comparisons at aggregate level among nations are affected by similar distortions. For example: if we take two universities or two countries (we shall call them A and B), that invest only in two areas (let us say Mathematical and Chemical Sciences), with B investing 50% more in total than A, of which 70% in Mathematical Sciences, in contrast to A, which invests the same percentage in Chemical Sciences, a simple aggregate calculation of the publications (and, probably, of citations) would put A ahead of B in terms of the contribution made to the scientific knowledge base.

Comparisons between universities or nations within the same disciplinary area (Adams, Table 2) would not be immune to the same problem of distortion due to the different levels of fertility of the individual scientific disciplines encompassed by the area, as we will show later.

For each of the eight disciplinary areas taken into consideration, we have split the researchers and relevant publications into their appropriate scientific disciplines and measured the intensity of publication for each SD. The statistics of the distribution of the intensity values are given in Table 2. The differences in publication intensity among the various SDs within the same DA are certainly not negligible, and are in fact even more marked than those among different DAs. In Industrial and Information Engineering, for example, the level of publication intensity in the most fertile SD is a full 39 times higher than that in the least fertile SD, and in Medicine it is 23 times higher.

**Table 2 - Descriptive statistics of publication intensity distributions across scientific disciplines in each disciplinary area, average 2001-2003**

| Disciplinary Area | Number of SDs | Min. | Max | Mean | Median | St. Dev. | Variation coeff. |
|---|---|---|---|---|---|---|---|

| | | | | | | |
|---|---|---|---|---|---|---|
| 1 – Mathematical sciences | 10 | 0.085 | 0.506 | 0.316 | 0.317 | 0.110 | 0.348 |
| 2 – Physical sciences | 8 | 0.205 | 1.699 | 1.046 | 1.001 | 0.498 | 0.476 |
| 3 –Chemical sciences | 12 | 0.742 | 2.143 | 1.322 | 1.394 | 0.419 | 0.317 |
| 4 – Earth sciences | 12 | 0.127 | 0.922 | 0.499 | 0.452 | 0.290 | 0.582 |
| 5 – Biological sciences | 19 | 0.205 | 1.379 | 0.813 | 0.858 | 0.327 | 0.402 |
| 6 – Medical sciences | 50 | 0.086 | 1.978 | 0.758 | 0.724 | 0.447 | 0.589 |
| 7 – Agricultural and veterinary sciences | 30 | 0.033 | 0.657 | 0.363 | 0.339 | 0.189 | 0.521 |
| 8 – Industrial and information engineering | 42 | 0.030 | 1.172 | 0.468 | 0.309 | 0.355 | 0.758 |

To give some idea of what may be the extent of the margin of error in those scientific performance rankings that do not take account of the SDs' different levels of fertility, we conducted two parallel evaluation exercises.

In the first of these exercises, we calculated the publication intensity of each university in each DA by simply dividing the total number of publications by the total number of researchers at the given university working in the given area. The second exercise replicated the first but focused on scientific disciplines. We then carried out an area-level aggregation through normalisation (with respect to the average scientific discipline publication intensity in all of the universities) and a weighing-up (with respect to the number of researchers in each discipline in relation to the area total)[11]. Subsequently, we compared the two rank orders of the universities using the two methodologies. Table 3 shows some statistics on position variation within the rankings. Both the number of universities changing position and the maximum and average variations in position in the two rankings attest to the significant distorting effect that bibliometric analyses at aggregate level can have. In our original, application-based study, we used a more complex indicator, which includes the quality of the publication (based on the number of citations) and the degree of ownership of the publication (based on the number of co-authors). The maximum and average variations were higher still.

**Table 3 - Comparisons of rank orders of universities per scientific discipline and disciplinary area based on publication intensity, 2001-2003**

| Disciplinary Area | Number of SDs | Number of Variations | Max Variation | Average Variation | Median |
|---|---|---|---|---|---|
| 1 – Mathematical sciences | 10 | 47 (out of 53) | 11 | 2.9 | 2 |
| 2 – Physical sciences | 8 | 53 (out of 55) | 20 | 4.2 | 3 |
| 3 –Chemical sciences | 12 | 48 (out of 57) | 16 | 4.0 | 3 |
| 4 – Earth sciences | 12 | 41 (out of 44) | 22 | 6.9 | 6 |
| 5 – Biological sciences | 19 | 52 (out of 58) | 35 | 6.0 | 3.5 |
| 6 – Medical sciences | 50 | 44 (out of 52) | 44 | 6.5 | 3 |
| 7 – Agricultural and veterinary sciences | 30 | 33 (out of 34) | 21 | 5.8 | 4 |
| 8 – Industrial and information engineering | 42 | 44 (out of 50) | 30 | 5.6 | 4 |

The distortions in the rank orders on quantity and quality of science in different nations become more acute in direct proportion to the extent of the differences in scientific specialisations across different nations. These distortions are increased when efficiency comparisons are made on the basis of indicators such as publications or citations per researcher and per HERD (higher education funding of R&D). In addition to the distortions noted above, there are also other distortions, which are due to the failure to differentiate between public sector and business outputs and inputs (see King, Fig. 5). As an example, let us see what happens to the research efficiency values of the different nations when they are measured with and without the distinction between public and private sector. To make calculation and comparison easier, we will use the simple value of publications per researcher[12] as our efficiency indicator. Using the latest available data on publications (dating mostly from 2003) and researchers (dating from 2002)[13], we calculated the value of the ratio for each G7 member state and the average of the EU-25 (our comparator group) in the two different ways (see Table 4).

By not distinguishing between public and private, and by dividing the total number of publications by the total number of researchers in a given country – i.e. in the same way as King – Italy would come out, by quite some distance, as the most efficient nation in terms of research activities, with a ratio that is one-third higher than that of the UK, double that of France, Germany and the EU-25, and two-and-a-half times that of the USA. Even if we consider as negligible the effect of scientific specialisation on the numerator, Italy's apparently brilliant performance – much as we might wish it otherwise – is, in fact, the result of the heavily distorting effect introduced by the failure to distinguish between public and private. The percentage of public sector researchers as part of the total is far higher in Italy than it is in the other countries within the comparator group (see Table 4), and, moreover, public researchers have a tendency to publish far more than private researchers – these factors account for the inaccuracy of the Italian performance value. A comparison of the number of scientific publications per researcher in the public sector across the various countries is made more complicated by the absence of data on the split between public and private sector publications. A less-than-perfect solution – which is, nevertheless, adopted by economists across the board – is based on the assumption that the contribution of the private researcher is zero. In overall terms, this is an acceptable approximation, but it becomes unacceptable as soon as comparisons are made between data from countries that have a very different public/private researcher ratio, as those in the comparator group do. By patiently and methodically indexing the affiliations of the authors, we have succeeded in distinguishing, to a relatively high degree of accuracy, the papers published by public research organisations and private companies in Italy. Since we knew the number of private researchers in Italy, we were able to calculate the average publication intensity of the private researchers, and this value was assumed to be constant across all of the countries of the comparator group[14]. By multiplying this by the number of private researchers in each country, we arrived at an estimate of the number of private publications. By subtracting this figure from the total number of publications, we came up with a figure for the number of public-sector publications. In this way, we could eliminate distortion due to the different splits between public and private researchers in the various countries, allowing us then to calculate, albeit approximately, the average annual scientific product of each national public researcher. The average number of publications per Italian public researcher – 0.82 – is now aligned with the average figures in Britain (0.86) and the USA (0.81), and remains higher than those of the other countries. Apart from Japan, the approach we have outlined changes the relative position of every member of the comparator group.

**Table 4 - Comparisons of national scientific performances with and without distinction between public and private sector, 2003 or latest available data**

| Indicator | I | F | D | UK | USA | J | C | EU-25 |
|---|---|---|---|---|---|---|---|---|
| Publications per researcher | 0.49 | 0.25 | 0.24 | 0.36 | 0.18 | 0.11 | 0.21 | 0.25 |
| Rank order | 1 | 3 | 5 | 2 | 7 | 8 | 6 | 3 |
| Public researchers as a percentage of the total | 59 | 45 | 40 | 40 | 19 | 31 | 38 | 50 |
| Publications per researcher in the public sector | 0.82 | 0.51 | 0.55 | 0.86 | 0.81 | 0.32 | 0.52 | 0.49 |
| Rank order | 2 | 6 | 4 | 1 | 3 | 8 | 5 | 7 |

**Discussion**

To cite one example: in their 2006 paper, Dosi et al.[15] object to the so-called 'European Paradox', which is the assertion that EU countries play a leading global role in terms of top-level scientific output, but lag quite far behind in their ability to convert this strength into wealth-generating innovations. They argue that Europe's weaknesses reside both in its system of scientific research and in its relatively weak industry. They base their thesis on the relative weakness of EU scientific research as defined, for the most part, by evidence submitted by King. The policy changes suggested by the authors include: much less emphasis on 'networking', 'interactions with the local environment', or 'attention to user needs' – current obsessions of European policy makers – and much more emphasis on policy measures aimed at strengthening 'frontier' research and, at the opposite end, at strengthening European corporate input. Dosi et al. are probably right when they state that the European Paradox does not appear in the data; but they are probably wrong when they use aggregate measurements to refute it. The upshot of this is that their proposals for research policy are highly questionable. Scientometrists should, then, be very cautious in presenting evidence of analyses that are affected by technical or methodological limits that significantly distort the results, since failure to apply the necessary degree of caution could have important and very negative consequences if the results are used by scholars in other disciplines or by policy decision-makers. Comparative bibliometric analyses cannot fail to recognise the range of levels of scientific fertility, not only within a given major disciplinary area but also within the different scientific disciplines encompassed by the same area.

Is it, then, possible to measure the scientific strength of one country compared to another? Perhaps the most direct method would be to compare the relative levels of investment in the different scientific disciplines or, as a compromise solution, to split the research personnel into their respective disciplines. But this data is not easily accessible[16] and, in any case, comparisons should be made on a shared basis of disciplinary classification that, where it currently exists, is certainly different from country to country. The only path that we feel can be followed, given the current state of affairs, is to refer back to the 168 subject categories through which the SCI indexes scientific journals, on condition that the temptation to provide aggregate comparative measurements is resisted.

The evidence that emerges from bibliometric analyses is, then, representative of the public scientific infrastructure of the country in question. Next to nothing, or nothing at all, emerges about private research, which in certain countries (UK and Germany) equates to 70% of total research and in other countries (USA and Japan) equates to an even higher percentage. For this reason, rather than talking about the 'scientific wealth of nations', it would be more accurate to talk about the 'public scientific potential of a nation or region'. If an adequate public scientific base is unanimously recognised as an essential condition for the sustainable competitiveness of the country's national economic system, then it is far from sufficient to actually achieve that sort of competitiveness. The capacity of the productive system to exploit that base is equally essential. National economic systems with excellent public scientific infrastructures but weak systems of public/private technology transfer could end up helping foreign economies more than their own, due to the ease with which cross-border migration of knowledge can occur through publications and the mobility of researchers. The capacity of the American system to attract the best minds, along with the consolidated tradition of collaboration between universities and industry, places the USA at an advantage over many EU countries in terms of its ability to make the most of the excellence of both domestic and foreign public scientific bases. But, as May said, "that's another story"… and a vital one, we would add.

[1] King, D.A.. *Nature* **430**, 311-316 (2004).

[2] May, R.M. *Science* **275**, 793-76 (1997).

[3] Van Raan, A.F.J. *Scientometrics* **62** (1), 133-143 (2005).

[4] Not to mention the difference in publication intensity between basic and applied research.

[5] Abbott and Doucouliagos, 2003 (*Economics of Education Review*, **22** (1), 89-97) and Worthington and Lee, 2005 (*University of Wollongong*, *working paper*) about Australian academies; Flegg et al., 2004 (*Education Economics*, **12** (3), 231-249) e Athanassopoulos and Shale, 1997 (*Education Economics*, **5**, 117–134) about UK universities; Baek, 2006 (*"New Frontiers In Evaluation" Conference*, Vienna) about US universities.

[6] SCI, Science Citation Index, indexes major international journals in science and engineering.

[7] Italian Ministry of Universities and Research.

[8] Scientific articles in arts cannot be considered fully representative of overall research outputs.

[9] The number of citations is correlated to the number of publications

[10] Adams, J. *Nature* **396**, 615-618 (1998).

[11] Publication intensity in disciplinary area *j* of university *k* equals:

$$\Theta_{Tot-k}(j) = \frac{\sum_{i=1}^{n_j} PQCn_{ik} Add_{ik}}{\sum_{i=1}^{n_j} Add_{ik}}, \text{ where}$$

$\Theta_{Tot-k}(j)$ = publication intensity in disciplinary area *j* of university *k*;

$PQCn_{ik}$ = publication intensity of university *k* in scientific discipline *i* (within area *j*), normalized to the average scientific discipline publication intensity in all of the universities;

$Add_{ik}$ = number of research staff associated with scientific discipline *i* (within area *j*) of university *k*;

$n_j$ = number of scientific disciplines encompassed in area *j*

[12] Because research activities are multi-input and multi output, efficiency measures should take into account these characteristics. More appropriate parametric and non parametric techniques to measure research efficiency have been developed and applied, and can be found in the relevant literature.

[13] OECD, Main Science and Technology Indicators 2006/1; and EC, Key Figures 2005.

[14] Not knowing the number of public sector publications of the other countries of the comparator group.

[15] Dosi, G., Llerema P., Sylos Labini M., *Research Policy* **35**, 1450-1464 (2006).

[16] In Italy, for example, differently from universities, researchers of government research laboratories are not grouped into scientific disciplines.